\documentclass[twocolumn,10pt]{IEEEtran}
\usepackage{ifpdf, flushend}

\ifCLASSINFOpdf
  \usepackage[pdftex]{graphicx}
  \graphicspath{{../pdf/}{../jpeg/}}
  \DeclareGraphicsExtensions{.pdf,.jpeg,.png}
\else
  \usepackage[dvips]{graphicx}
  \graphicspath{{../eps/}}
  \DeclareGraphicsExtensions{.eps}
\fi
\usepackage[cmex10]{amsmath}
\usepackage {amssymb}
\usepackage{algorithmic}
\usepackage{array}
\usepackage{mdwmath}
\usepackage{mdwtab}
\usepackage{eqparbox}
\usepackage{url}
\usepackage{hyperref}
\usepackage{algorithm}
\usepackage{algorithmic}

\newcommand{\beq}{\begin{equation}}
\newcommand{\eeq}{\end{equation}}
\newcommand{\beqn}{\begin{eqnarray}}
\newcommand{\eeqn}{\end{eqnarray}}
\newcommand{\beqno}{\begin{eqnarray*}}
\newcommand{\eeqno}{\end{eqnarray*}}
\newcommand{\bma}{\begin{displaymath}}
\newcommand{\ema}{\end{displaymath}}
\newcommand{\bnu}{\begin{enumerate}}
\newcommand{\enu}{\end{enumerate}}
\newcommand{\bce}{\begin{center}}
\newcommand{\ece}{\end{center}}
\newcommand{\btb}{\begin{tabular}}
\newcommand{\etb}{\end{tabular}}

\begin{document}

%
\title{Multi-Channel MAC Protocol for Full-Duplex Cognitive Radio Networks with Optimized Access Control and Load Balancing}

\author{\IEEEauthorblockN{Le Thanh Tan and Long Bao Le}  
\thanks{The authors are with INRS-EMT, University of Quebec,  Montr\'{e}al, Qu\'{e}bec, Canada. 
Emails: \{lethanh,long.le\}@emt.inrs.ca.}}

\maketitle

\begin{abstract}
\boldmath
In this paper, we propose a multi-channel full-duplex  Medium Access Control (MAC) protocol for cognitive radio networks (MFDC--MAC). 
Our design exploits the fact that full-duplex (FD) secondary users (SUs) can perform spectrum sensing and access simultaneously, and we
employ the randomized dynamic channel selection for load balancing among channels and the standard backoff mechanism 
for contention resolution on each available channel. Then, we develop a mathematical model to analyze the throughput 
performance of the proposed MFDC--MAC protocol. Furthermore, we study the protocol configuration optimization to maximize
the network throughput where we show that this optimization can be performed in two steps, namely optimization of access
and transmission parameters on each channel and optimization of channel selection probabilities of the users. 
Such optimization aims at achieving efficient self-interference management for FD transceivers, sensing overhead control, 
and load balancing among the channels. Numerical results demonstrate the impacts of different protocol parameters and
the importance of parameter optimization on the throughput performance as well as the significant performance gain 
of the proposed design compared to traditional design.
\end{abstract}

\IEEEpeerreviewmaketitle

\section{Introduction}

Design of MAC protocols for efficient sharing of white spaces and appropriate protection of transmissions from primary users (PUs) on 
licensed frequency in cognitive radio networks (CRNs) is an important research topic.
In the traditional design and analysis of a half-duplex (HD) MAC protocol \cite{Cor07,Park11,Liang08, Tan11, Tan12},
 SUs typically employ a two-stage sensing/access procedure due to the HD constraint \cite{Liang08, Tan11, Tan12}. 
This constraint also requires SUs be synchronized during the spectrum sensing stage, which could be difficult to achieve in practice.  
Moreover, sophisticated design and parameter configuration of cognitive MAC protocols can result in significant performance enhancement
 while appropriately protecting SUs \cite{Tan11, Tan12}. Furthermore, different multi-channel cognitive MAC protocols were proposed considering 
either different spectrum sensing and access methods \cite{Cor07,Park11}.

By employing the advanced FD transceiver, each SU can transmit and receive data simultaneously on the same frequency band \cite{Duarte12}.  
Practical FD transceivers, however, suffer from self-interference, which is caused by power leakage from the transmitter to the receiver.
The self-interference may indeed lead to serious communication performance degradation of FD wireless systems.
Employment of FD transceivers for more efficient spectrum access design in cognitive radio networks has been very under explored in
the literature. The cognitive MAC design in one recent work \cite{Cheng14} allows SUs to perform sensing and transmission simultaneously;  however,  
the work \cite{Cheng14} assumes simultaneous spectrum access of the SU and PU networks. 
This design is, therefore, not applicable to the hierarchical spectrum access in the CRNs where PUs should have 
higher spectrum access priority compared to SUs.

In this paper, we propose a novel MFDC--MAC protocol that allows concurrent spectrum sensing
and transmission on each channel as well as efficient access and load balancing among the channels. 
In our design, each SU adopts the randomized channel selection to choose its channel, which is slowly
updated over time for load balancing. 
Moreover, SUs employ the standard $p$-persistent CSMA mechanism for contention resolution on the selected channel, 
and the winning SU follows a two-stage procedure for spectrum sensing and access. Specifically,
the winning SU performs simultaneous sensing and transmission during the first stage and transmission only in the second stage.
This design enables appropriate protection of PUs and efficient exploitation of white spaces on all the channels.

We develop a mathematical model for throughput performance analysis of the proposed MFDC-MAC protocol considering the imperfect sensing 
and self-interference effects. Moreover, we study the optimal configuration of different protocol parameters for spectrum sensing, access,
and load balancing (i.e., channel access probabilities) to achieve the maximum throughput. 
Extensive numerical results are then presented to illustrate the impacts of different 
protocol parameters on the throughput performance and the significant throughput gains of the proposed MFD-MAC protocol
with respect to conventional designs.

The remaining of this paper is organized as follows. Section II describes the system and PU activity models. MAC protocol design and
throughput analysis are performed in Section III. We discuss the protocol optimization in Section IV. Section V demonstrates 
numerical results followed by concluding remarks in Section ~VI.

\section{System and PU Activity Models}
\label{SystemModel}

\subsection{System Model}
\label{System}

We consider a network setting where $N$ pairs of SUs opportunistically exploit white spaces on $M$ frequency channels for data transmission. 
We assume that each SU is equipped with one full-duplex transceiver, which can perform sensing and transmission simultaneously.
However, any SU suffers from self-interference from its transmission during sensing (i.e., transmitted signals are leaked into
the received signal). At channel $j$, we denote $I_j$ as the average self-interference power, which is assumed to be modeled as
 $I_j = \zeta \left(P_{{\sf sen},j}\right)^{\xi}$ \cite{Duarte12} where $P_{{\sf sen},j}$ is the SU transmit power, 
$\zeta$ and $\xi$ ($0 \leq \xi \leq 1$) are
predetermined coefficients which capture the self-interference cancellation quality.
We design an asynchronous MAC protocol where no synchronization is required between SUs and PUs as well as among SUs.
We assume that different pairs of SUs can overhear transmissions from the others (i.e., a collocated network). 
In the following, we refer to pair $i$ of SUs simply as SU $i$.

\subsection{Primary User Activity}
\label{PUAM}

We assume that the PU's idle/busy status follows two independent and identical distribution processes. In particular, each channel is available and busy 
for the secondary access if the PU is in the idle and busy states, respectively. Let $\mathcal{H}_0$ and $\mathcal{H}_1$ denote the events 
that the PU is idle and active, respectively. To protect the PU, we assume that SUs must stop their transmission and evacuate from the channel 
within the maximum delay of $T_{\sf eva}$, which is referred to as channel evacuation time. 

Let $\tau_{\sf ac}^j$ and $\tau_{\sf id}^j$ denote the random variables which represent the durations of channel active and idle states on channel $j$, respectively.
We assume that $\tau_{\sf ac}^j$ and $\tau_{\sf id}^j$ are larger than $T_{\sf eva}$ with high probability.
We denote probability density functions of $\tau_{\sf ac}^j$ and $\tau_{\sf id}^j$ as $f_{\tau_{\sf ac}^j}\left(t\right)$ and $f_{\tau_{\sf id}^j}\left(t\right)$, respectively.  
In addition, let  $\mathcal{P}\left(\mathcal{H}_0^j\right) = \frac{{\bar \tau}_{\sf id}^j}{{\bar \tau}_{\sf id}^j+{\bar \tau}_{\sf ac}^j}$ and  $\mathcal{P}\left( \mathcal{H}_1^j \right) = 
1 - \mathcal{P}\left(\mathcal{H}_0^j\right)$ present the probabilities that the channel is available and busy, respectively.

\section{Multi-Channel  Full-Duplex Cognitive MAC Protocol}

In this section, we describe our proposed MFDC-MAC protocol and conduct its throughput
analysis considering imperfect sensing and self-interference of the FD transceiver.

\subsection{MFDC--MAC Protocol Design}

\begin{figure}[!t]
\centering
\includegraphics[width=90mm]{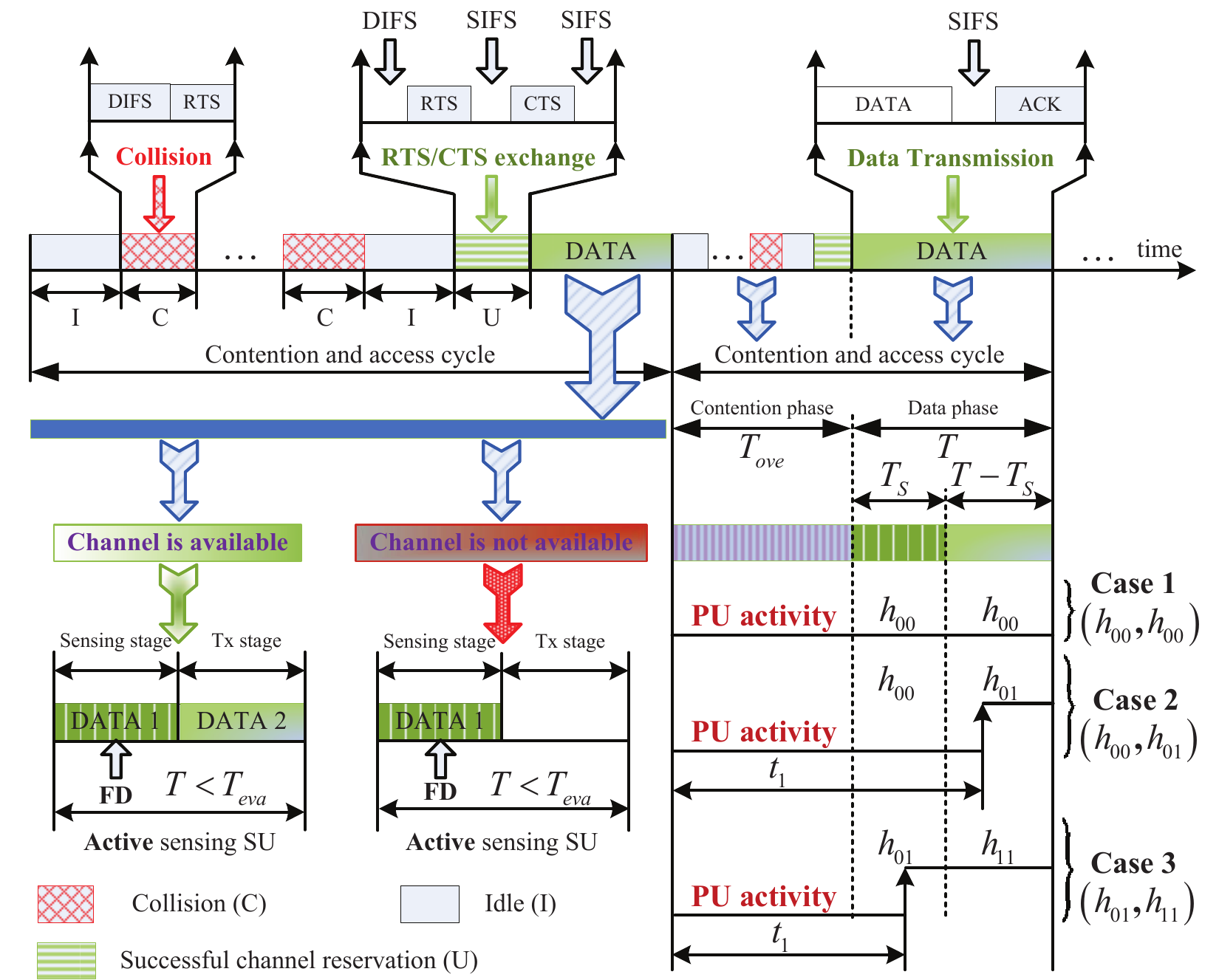}
\caption{Timing diagram of the proposed full-duplex MAC protocol.}
\label{Sentime_FDMAC_SF}
\end{figure}

In our MFDC-MAC protocol, each SU randomly selects one channel by using a randomized channel 
selection mechanism where channel $j$ is selected with probability $p_j^{\sf sec}$.
In this paper, we consider the general heterogeneous scenario where the statistical parameters 
$\tau_{\sf ac}^j$ and $\tau_{\sf id}^j$ of different channel $j$ can be different.
This channel selection is repeated once after a predetermined long period, which is an order
of magnitude larger than the average contention/access time to transmit one data frame (packet) on each channel
(e.g., every $\mathcal{K}_{\sf max}$ data frames). 

After channel selection, each SU employs the following single-channel contention, spectrum sensing,
and transmission to exploit the white space.  
Specifically, SUs choosing the same channel $j$ is assumed to employ the $p$-persistent CSMA principle \cite{Cali00} for contention resolution
where each SU attempts to capture the channel with a probability $p$ after the channel is sensed to be idle during the standard DIFS interval
 (DCF Interframe Space). If a particular SU decides not to transmit (with probability of $1-p$), it will carrier sense the channel and 
attempt to transmit again in the next slot with probability $p$.

To complete the reservation, the four-way handshake with Request-to-Send/Clear-to-Send (RTS/CST) exchanges \cite{Cali00} is employed to reserve the available channel for transmission in the next phase. After each successful transmission of duration $T_j$, an acknowledgment (ACK) from the SU's receiver is transmitted to its corresponding transmitter to notify the successful reception of a packet.
Furthermore, the standard small interval, namely SIFS (Short Interframe Space), is used before the transmissions of CTS, ACK, and data frame as in the
standard 802.11 MAC protocol \cite{Cali00}.

Then, the data phase after the channel contention phase comprises two stages where the winning SU performs
concurrent sensing and transmission in the first stage with duration $T_{S,j}$ (called FD sensing stage) and 
transmission only in the second stage with duration $T_j-T_{S,j}$ (called transmission stage).
Here, the SU exploits the FD communication capability of its transceiver to realize concurrent sensing and transmission 
the first stage   where the sensing outcome at the end of this stage (i.e., an idle or active channel status) determines its further actions as described in the following.
If the sensing outcome indicates an available channel then the SU transmits data in the second stage; otherwise, it remains silent for the remaining period of the data phase with duration $T_j-T_{S,j}$.

We assume that the duration of the SU's data phase $T_j$ is smaller than the channel evacuation time $T_{\sf eva}$ so timely evacuation 
from the busy channel can be realized. Therefore, our design allows to protect the PU with evacuation delay at most $T_j$ if the carrier 
sensing before the contention phase and the spectrum sensing in the data phase are perfect. Furthermore, we assume that the SU transmits 
at power levels $P_{{\sf sen},j}$ and $P_{\sf dat}$ during
the FD sensing and transmission stages, respectively where the transmit power $P_{{\sf sen},j}$ will be chosen to effectively mitigate the self-interference and achieve good sensing-throughput tradeoff. The timing diagram of the proposed MFDC--MAC protocol is illustrated in 
Fig.~\ref{Sentime_FDMAC_SF}.

\subsection{Throughput Analysis}

We now analyze the saturation throughput. Recall that for each channel $j$, the PU is active and idle with the
 corresponding pdfs of $\tau_{\sf ac}^j$ and $\tau_{\sf id}^j$ are $f_{\tau_{\sf ac}^j}\left(\tau_{\sf ac}^j\right)$ 
and $f_{\tau_{\sf id}^j}\left(\tau_{\sf id}^j\right)$. Moreover, we assume that the received PU's signal power at the SU's receiver for channel $j$ is $P_p^j$. 
The throughput is a function of following parameters: probability of transmission $p$, sensing time $T_{S,j}$, frame length $T_j$, and SU transmit 
power $P_{{\sf sen},j}$ for each channel $j$. For brevity, we ignore the dependence of throughput on $p$ and $T_j$ in the following.

To calculate the throughput of the MFDC--MAC protocol, denoted as $\mathcal{NT}$, we consider all
possible cases where each case is represented by the corresponding sets of users selecting different channels 
(each set of users is for one channel). We now define the set $\Omega = \left\{\! \omega_k \!\!=\!\! \left\{n_{k,1}, \ldots, n_{k,j}, \ldots, 
n_{k,M}\!\right\}\!\!: \sum_{j=1}^M \!{n_{k,j}} \!= \!N\right\}$ where its $k$-th element $\omega_k$
has its components $n_{k,j}$ representing the number of users who select channel $j$.
The probability for the set $\omega_k$ is $\prod_{ j =1}^{M} \! \left(p_j^{\sf sec}\right)^{n_{k,j}}$.
Then, the network throughput can be expressed as
\beqn
\mathcal{NT} \!\!\left( \vec{\mathcal{P}}^{\sf sec}, \vec{T}_S, \vec{P}_{\sf sen}\right)\!\! = \!\!\sum_{k=1}^{\left|\Omega\right|} \!\! 
\left(\!\!\!\!{\begin{array}{*{20}{c}} N  \\ \left\{ n_{k,j} \right\}  \\ \end{array}}
\!\!\!\! \right) \! \prod_{ j =1}^{M} 
\!\! \left(p_j^{\sf sec}\right)^{n_{k,j}} \label{NT_MGM_1} \nonumber \\
 \sum_{j=1}^M \mathcal{NT}_j \left(T_{S,j}, P_{{\sf sen},j}\left|n_{k,j}\right.\right) \mathcal{I} \left(n_{k,j} > 0\right) \label{NT_MGM_2},
\eeqn
where $\mathcal{NT}_j \left(T_{S,j}, P_{{\sf sen},j}\left|n_{k,j}\right.\right) \mathcal{I} \left(n_{k,j} > 0\right)$
represents the throughput contributed by channel $j$ given that $n_{k,j}$ users select this channel; and
$\mathcal{I}(.)$ denotes the indicator function. Moreover, 
$\left(\!\!\!\!{\begin{array}{*{20}{c}} N  \\ \left\{ n_{k,j} \right\}  \\ \end{array}} \!\!\!\!\right)$ is the multinomial coefficient which is defined as $\left(\!\!\!\!{\begin{array}{*{20}{c}} N  \\ \left\{ n_{k,j} \right\}  \\ \end{array}} \!\!\!\!\right) = \left(\!\!\!\!{\begin{array}{*{20}{c}} N  \\ n_{k,1}, n_{k,2}, \ldots, n_{k,M}   \\ \end{array}} \!\!\!\!\right) = \frac{N}{n_{k,1}! n_{k,2}! \ldots n_{k,M} !}$.

Furthermore, the throughput of channel $j$ $\mathcal{NT}_j \left( T_{S,j}, P_{{\sf sen},j}\left|n_{k,j}\right.\right)$ can be calculated as 
\beqn
\mathcal{NT}_j \left(T_{S,j}, P_{{\sf sen},j}\left|n_{k,j}\right.\right) = \frac{B_j}{T_{{\sf{ove}},j}+T_j}.
\eeqn
where $T_{{\sf{ove}},j}$ represents the time overhead required for one successful channel reservation on channel $j$ (i.e., successful RTS/CTS exchanges), 
$B_j$ (bits/Hz) denotes the average number of  bits transmitted per one unit of system bandwidth for one contention/access (CA) cycle on channel $j$.
To complete the throughput analysis, we derive the quantities $T_{\sf{ove},j}$ and $B_j$, which are conducted in the following.

\subsubsection{Derivation of $T_{\sf{ove},j}$}

The average time overhead for one successful channel reservation can be written as
\beqn \label{tover}
T_{\sf{ove},j} = {\overline T}_{{\sf cont},j} + 2SIFS + 2PD + ACK,
\eeqn
where $ACK$ is the length of an ACK message, $SIFS$ is the length of a short interframe space, and $PD$ is the propagation delay
where $PD$ is usually small compared to the slot size $\sigma$, and ${\overline T}_{{\sf cont},j}$ denotes the average time overhead
due to idle periods, collisions, and successful transmissions of RTS/CTS messages in one CA cycle.
To calculate ${\overline T}_{{\sf cont},j}$, we define some further parameters as follows.
Denote $T_{\sf coll}$ as the duration of the collision and
$T_{\sf succ}$ as the required time for successful RTS/CTS transmission. These quantities can be calculated as follows \cite{Cali00}:
\beqn
\label{TCTSTI}
\left\{ {\begin{array}{*{20}{c}}
   T_{\sf succ} = DIFS + RTS + SIFS + CTS + 2PD \hfill  \\
   T_{\sf coll} = DIFS + RTS + PD, \hfill  \\
\end{array}} \right.
\eeqn
where $DIFS$ is the length of a distributed interframe space, $RTS$ and $CTS$ denote
the lengths of the RTS and CTS messages, respectively.

As being shown in Fig.~\ref{Sentime_FDMAC_SF}, there can be several idle periods and collisions before one successful channel reservation.
Let $T_{{\sf idle},j}^i$ denote the $i$-th idle duration between two consecutive RTS/CTS exchanges on channel $j$, which can be collisions or successful exchanges. 
Then, $T_{\sf idle}^i$ can be calculated based on its probability mass function (pmf),  which is derived as follows. 
In the following, all relevant quantities are defined in terms of the number of time slots.
With $n_{k,j}$ SUs joining the contention resolution on channel $j$, let $\mathcal{P}_{{\sf succ},j}$, $\mathcal{P}_{{\sf coll},j}$ and $\mathcal{P}_{{\sf idle},j}$ denote
 the probabilities that a particular generic slot corresponds to a successful transmission, a collision, and an idle slot, respectively. These 
probabilities can be calculated as follows:
\beqn
\mathcal{P}_{{\sf succ},j} &=& n_{k,j}p\left(1-p\right)^{n_{k,j}-1} \\
\mathcal{P}_{{\sf idle},j} &=&  \left(1-p\right)^{n_{k,j}} \hspace{1cm} \\
\mathcal{P}_{{\sf coll},j} &=&  1-\mathcal{P}_{{\sf succ},j}-\mathcal{P}_{{\sf idle},j},
\eeqn
where $p$ is the transmission probability of an SU in a generic slot. 
In general, the interval ${ T}_{{\sf cont},j}$, whose average value is  ${\overline T}_{{\sf cont},j}$ given in (\ref{tover}), consists of several intervals 
corresponding to idle periods, collisions, and one successful RTS/CTS transmission. 
Hence, this quantity can be expressed as 
\beqn
\label{T_cont}
{T}_{{\sf cont},j} = \sum_{i=1}^{N_{{\sf coll},j}} \left(T_{\sf coll}+ T_{{\sf idle},j}^i\right) + T_{{\sf idle},j}^{N_{{\sf coll},j}+1} + T_{\sf succ},
\eeqn
where $N_{{\sf coll},j}$ is the number of collisions before the successful RTS/CTS exchange on channel $j$ and 
$N_{{\sf coll},j}$ is a geometric random variable (RV) with parameter $1-\mathcal{P}_{{\sf coll},j}/\mathcal{\overline P}_{{\sf idle},j}$ where $\mathcal{\overline P}_{{\sf idle},j} = 1 - \mathcal{P}_{{\sf idle},j}$. 
Therefore, its pmf can be expressed as
\beqn
\label{N_c_cal}
 f_{X}^{N_{{\sf coll},j}} \left(x\right) = \left(\frac{\mathcal{P}_{{\sf coll},j}}{\mathcal{\overline P}_{{\sf idle},j}}\right)^{x} \left(1-\frac{\mathcal{P}_{{\sf coll},j}}{\mathcal{\overline P}_{{\sf idle},j}}\right), \: x = 0, 1, 2, \ldots
\eeqn
Also, $T_{{\sf idle},j}$ represents the number of consecutive idle slots on channel $j$, which is also a geometric RV with parameter $1-\mathcal{P}_{{\sf idle},j}$ with the following pmf
\beqn
\label{T_I_cal}
f_{X}^{T_{{\sf idle},j}} \left(x\right) = \left(\mathcal{P}_{{\sf idle},j}\right)^{x} \left(1-\mathcal{P}_{{\sf idle},j}\right), \: x = 0, 1, 2, \ldots
\eeqn
Therefore, ${\overline T}_{{\sf cont},j}$ (the average value of ${T}_{{\sf cont},j}$) can be written as follows \cite{Cali00}:
\beqn
{\overline T}_{{\sf cont},j}  = {\overline N}_{{\sf coll},j} T_{\sf coll} + {\overline T}_{{\sf idle},j} \left({\overline N}_{{\sf coll},j}+1\right) + T_{\sf succ} \label{T_contgeo},
\eeqn
where ${\overline T}_{{\sf idle},j}$ and ${\overline N}_{{\sf coll},j}$ can be calculated as
\beqn
{\overline T}_{{\sf idle},j} &=& \frac{\left(1-p\right)^{n_{k,j}}}{1-\left(1-p\right)^{n_{k,j}}} \\
{\overline N}_{{\sf coll},j} &=& \frac{1-\left(1-p\right)^{n_{k,j}}}{n_{k,j}p\left(1-p\right)^{n_{k,j}-1}}-1. 
\eeqn
These expressions are obtained by using the  pmfs of the corresponding RVs given in (\ref{N_c_cal}) and (\ref{T_I_cal}), respectively \cite{Cali00}.

\subsubsection{Derivation of $B_j$}

To calculate $B_j$, we consider all possible cases that capture the activities of SUs and status changes of the PU in the data phase of duration $T_j$.
Because the PU's activity is not synchronized with the SU's transmission, the PU can change its active/inactive status any time.
We assume that there can be at most one transition between the idle and active states of the PU during the interval $T_j$. 
This is consistent with the assumption on the slow status changes of the PU as described in Section~\ref{PUAM} since $T_j < T_{\sf eva}$. 
Furthermore, we assume that the carrier sensing of the MFDC-MAC protocol is perfect; therefore, the PU is idle at the beginning of the data phase.
Note that the PU may change its status during the SU's sensing or access stage, which requires us to consider different possible events in the data phase. 

We use $h_{kl}$ ($k, l \in \left\{ 0, 1 \right\}$) to represent events capturing status changes of the PU in the FD sensing stage and transmission stage
where $i$ = 0 and $i$ = 1 represent the idle and active states of the PU, respectively.
For example, if the PU is idle during the FD sensing stage and becomes active during the transmission stage, then we represent
this event as $\left(h_{00}, h_{01}\right)$ where sub-events $h_{00}$ and $h_{01}$ represent the status changes in the FD sensing
and transmission stages, respectively. Moreover, if the PU changes from the idle to the active state during
the FD sensing stage and remains active in the remaining of the data phase, then we represent this event as $\left(h_{01}, h_{11}\right)$

It can be verified that we must consider the following three cases with the corresponding status changes of the PU during the FDC-MAC data phase to analyze $\mathcal{B}_j$.
\begin{itemize}
\item \textbf{Case 1}: The PU is idle for the whole FDC-MAC data phase  (i.e., there is no PU's signal in both FD sensing and transmission stages) and we denote this
event as $\left(h_{00}, h_{00}\right)$. The average number of bits (in bits/Hz) transmitted during the data phase in this case is denoted as $B_{j,1}$.

\item \textbf{Case 2}: The PU is idle during the FD sensing stage but the PU changes from the idle to the active status in the transmission stage. 
We denote the event corresponding to this case as $\left(h_{00}, h_{01}\right)$ where $h_{00}$ and $h_{01}$ capture the sub-events in the FD sensing and transmission stages, respectively.
The average number of bits (in bits/Hz) transmitted during the data phase in this case is represented by $B_{j,2}$.

\item \textbf{Case 3}: The PU is first idle then becomes active during the FD sensing stage and it remains active during the whole transmission stage.
Similarly we denote this event as $\left(h_{01}, h_{11}\right)$ and the average number of bits (in bits/Hz) transmitted during the data phase in this case is denoted as $B_{j,3}$.
\end{itemize}

Then, we can calculate $\mathcal{B}_j$ as follows:
\beqn
B_j = B_{j,1} + B_{j,2} + B_{j,3}.
\eeqn
Theoretical derivation for $B_{j,1}$, $B_{j,2}$, and $B_{j,3}$ is given in the online technical report \cite{report} due to the space constraint.

\section{MFDC--MAC Protocol Configuration for Throughput Maximization}
\label{FDC_MAC_Configuration}

In this section, we  study the optimal configuration of the proposed MFDC--MAC protocol to achieve the maximum secondary
 throughput while satisfactorily protecting the PU. 

\subsection{Problem Formulation}
\label{TputOpt}

We are interested in determining suitable configuration for  $\vec{\mathcal{P}}^{\sf sec}$, $\vec{T}_S$,  and $\vec{P}_{\sf sen}$ to maximize the secondary throughput, $\mathcal{NT} \left( \vec{\mathcal{P}}^{\sf sec}, \vec{T}_S, \vec{P}_{\sf sen}\right)$.
Suppose that the PU requires that the average detection probability be at least $\overline{\mathcal{P}}_{d}^j$.
Then, the throughput maximization problem can be stated as follows:

\vspace{0.0cm}
\noindent
\textbf{P1:} 
\vspace{0.1cm}
\begin{equation}
\label{EQN_Problem1}
\begin{array}{l}
 {\mathop {\max }\limits_{p, \vec{\mathcal{P}}^{\sf sec}, \vec{T}_S, \vec{P}_{\sf sen}}} \quad {\mathcal{NT}} \left(\vec{\mathcal{P}}^{\sf sec}, \vec{T}_S, \vec{P}_{\sf sen}\right)  \\ 
 \mbox{s.t.}\,\,\,\, \hat{\mathcal{P}}_{d}^j\left(\varepsilon^j,T_{S,j}\right) \geq \mathcal{\overline P}_{d}^j, \quad 0 \leq P_{{\sf sen},j} \leq P_{\sf max},  \\
 \quad 0 \leq T_{S,j} \leq T_j, \quad 0 \leq p^{\sf sec}_j \leq 1, \quad \sum_{j=1}^M p^{\sf sec}_j = 1 \\
 \end{array}\!\!
\end{equation}
where $p^{\sf sec}_j$ is the probability of channel selection ($\vec{\mathcal{P}}^{\sf sec} = \left\{p^{\sf sec}_j\right\}$), $P_{{\sf sen},j}$ is the SU's transmit power on channel $j$ and $P_{\sf max}$ is the maximum power for SUs, and $T_{S,j}$ is upper bounded by $T_j$.
In fact, the first constraint on $\hat{\mathcal{P}}_{d}^j\left(\varepsilon^j,T_{S,j}\right)$ implies that the spectrum sensing should be sufficiently reliable to protect the PU. Moreover, the SU's transmit power $P_{{\sf sen},j}$ must be appropriately set to achieve good tradeoff between the network throughput and 
self-interference mitigation. 

To solve problem (\ref{EQN_Problem1}), we propose the two-step approach where we solve the following two subproblems \textbf{P2} and \textbf{P3} 
in the two steps, respectively.
In the first stage, we optimize the parameters for each individual channel $j$ and $n_{k,j}$ contending SUs on this channel to achieve maximum throughput of
each channel $j$, i.e., $\mathcal{NT}_j \left(T_{S,j}, P_{{\sf sen},j}\left|n_{k,j}\right.\right)$.
This problem can be presented as

\vspace{0.0cm}
\noindent
\textbf{P2:} 
\vspace{0.1cm}
\begin{equation}
\label{EQN_Problem2}
\begin{array}{l}
 {\mathop {\max }\limits_{T_{S,j}, P_{{\sf sen},j}}} \quad \mathcal{NT}_j \left(T_{S,j}, P_{{\sf sen},j}\left|n_{k,j}\right.\right)   \\ 
 \mbox{s.t.}\,\,\,\, \hat{\mathcal{P}}_{d}^j\left(\varepsilon^j,T_{S,j}\right) \geq \mathcal{\overline P}_{d}^j, \quad 0 \leq P_{{\sf sen},j} \leq P_{\sf max},  \\
 \quad \quad 0 \leq T_{S,j} \leq T_j\\
 \end{array}\!\!
\end{equation}
After solving problem \textbf{P2} with optimal results $\mathcal{NT}_j^* \left( T_{S,j}^*(n_{k,j}), P_{{\sf sen},j}^*(n_{k,j})\left|n_{k,j}\right.\right)$ 
for all channels $j$ and possible cases with different $n_{k,j}$ contending SUs. Then, the network throughput with given $\vec{T}_S^*, \vec{P}_{\sf sen}^*$
only depends on the channel selection probabilities in $\vec{\mathcal{P}}^{\sf sec}$.
Problem \textbf{P3} maximizes the throughput with respect to $\vec{\mathcal{P}}^{\sf sec}$ as

\vspace{0.0cm}
\noindent
\textbf{P3:} 
\vspace{0.1cm}
\begin{equation}
\label{EQN_Problem3}
\begin{array}{l}
 {\mathop {\max }\limits_{\vec{\mathcal{P}}^{\sf sec}}} \quad {\mathcal{NT}} \left(\vec{\mathcal{P}}^{\sf sec}\right)  \\ 
 \mbox{s.t.}\,\,\,\, \quad 0 \leq p^{\sf sec}_j \leq 1 \quad \sum_{j=1}^M p^{\sf sec}_j = 1. \\
 \end{array}\!\!
\end{equation}
Here, ${\mathcal{NT}} \left(\vec{\mathcal{P}}^{\sf sec}\right)$ can be written as
\beqn
\mathcal{NT} \!\!\left(\vec{\mathcal{P}}^{\sf sec}\right)\!\! = \sum_{k=1}^{\left|\Omega\right|} 
\prod_{j=1}^M \! \left(p_j^{\sf sec}\right)^{n_{k,j}} \mathcal{B} \left(\left\{n_{k,j}\right\}\right)
\eeqn
where 
\beqn\label{EQN_B_constant}
\mathcal{B} \left(\omega_k\right) = \mathcal{B} \left(\left\{n_{k,j}\right\}\right)= \left({\begin{array}{*{20}{c}} N  \\ \left\{ n_{k,j} \right\}  \\ \end{array}} \right) \sum_{j =1}^M \mathcal{I} \left(n_{k,j} > 0\right) \times \nonumber\\
\mathcal{NT}_j^* \left( T_{S,j}^*(n_{k,j}), P_{{\sf sen},j}^*(n_{k,j})\left|n_{k,j}\right.\right).
\eeqn
Due to the decomposed structure of the throughput expression ${\mathcal{NT}} \left(\vec{\mathcal{P}}^{\sf sec}, \vec{T}_S, \vec{P}_{\sf sen}\right)$
in (\ref{NT_MGM_2}), it can be seen that the proposed two-step approach does not loose optimality.

\subsection{Configuration for MFDC--MAC Protocol}

\subsubsection{Configuration for Sensing and Access Stages}

We will solve problem \textbf{P2} in the following.
In the following analysis, we assume the exponential distribution for ${\tau}_{\sf ac}^j$ and ${\tau}_{\sf id}^j$ where ${\bar \tau}_{\sf ac}^j$ and ${\bar \tau}_{\sf id}$ denote the
corresponding average values of these active and idle intervals on channel $j$.
Specifically, let $f_{\tau_{\sf x}^j}\left(t\right)$ denote the pdf of $\tau_{\sf x}^j$ (${\sf x}$ represents ${\sf ac}$ or ${\sf id}$ in the pdf of $\tau_{\sf ac}^j$ or $\tau_{\sf id}^j$, respectively) 
then
\beqn
\label{pdf_tau_ac_id}
f_{\tau_{\sf x}^j}\left(t\right) =  \frac{1}{{\bar \tau}_{\sf x}^j} \exp(-\frac{t}{{\bar \tau}_{\sf x}^j}).
\eeqn
We assume a homogeneous case with same frame length $T_j$.
We are interested in determining suitable configuration for $T_{S,j}$, and $P_{{\sf sen},j}$ to maximize the secondary throughput,
$\mathcal{NT}_j \left( T_{S,j}, P_{{\sf sen},j}\left|n_{k,j}\right.\right)$.
To gain insights into the parameter configuration of the MFDC--MAC protocol, we first study the optimization with respect to the sensing time $T_{S,j}$ for 
a given $P_{{\sf sen},j}$.

For a fixed $T_{S,j}$, we would need to set the sensing detection threshold $\varepsilon^j$ 
so that  the detection probability constraint
is met with equality, i.e., $\mathcal{\hat P}_d^j \left(\varepsilon^j,T_{S,j}\right) = \mathcal{\overline P}_d^j$ as in \cite{Liang08, Tan11}. 
Since the detection probability is smaller in \textbf{Case 3} (i.e., the PU changes from the idle to 
active status during the FD sensing stage of duration $T_{S,j}$) compared to that in \textbf{Case 1} and \textbf{Case 2} (i.e.,
the PU remains idle during the FD sensing stage) considered in the previous section, we only need to consider \textbf{Case 3}
to maintain the detection probability constraint. The average probability of detection for the FD sensing in \textbf{Case 3} can be expressed as
\beqn
\label{P_average}
\mathcal{\hat P}_d^j = \!  \int_{0}^{T_{S,j}}  \mathcal{P}_d^{j,01}(t) \! f_{{\tau}_{\sf id}^j}\!\!\left(t\left|0 \leq t \leq T_{S,j}\right.\right) dt,
\eeqn
where $t$ denotes the duration from the beginning of the FD sensing stage to the instant when the PU changes to the active state, and
$f_{{\tau}_{\sf id}^j}\left(t\left|\mathcal{A}\right.\right)$ is the pdf of ${\tau}_{\sf id}^j$ conditioned on event $\mathcal{A}$
capturing the condition $0 \leq t \leq T_{S,j}$, which is given as
\beqn
\label{cond_pdf_tau_id}
f_{{\tau}_{\sf id}^j}\left(t\left|\mathcal{A}\right.\right) = \frac{f_{{\tau}_{\sf id}^j}\left(t\right)}{\Pr\left\{\mathcal{A}\right\}} = \frac{\frac{1}{{\bar \tau}_{\sf id}^j} \exp(-\frac{t}{{\bar \tau}_{\sf id}^j})}{1-\exp(-\frac{T_{S,j}}{{\bar \tau}_{\sf id}^j})}.
\eeqn
Note that $\mathcal{P}_d^{j,01}(t)$ is derived in Appendix \ref{CAL_P_F_P_D} and $f_{{\tau}_{\sf id}^j} \left(t\right)$ is given in (\ref{pdf_tau_ac_id}). 

We consider the following single-variable optimization problem for a given $P_{{\sf sen},j}$:
\begin{equation}
\label{NT_NonFrag_OPT_TS}
{\mathop {\max }_{0< T_{S,j} \leq T} } \quad {\mathcal{NT}_j} \! \left(T_{S,j}, P_{{\sf sen},j} \left|n_{k,j}\right. \right). 
\end{equation}
We characterize the properties of function $\mathcal{NT}_j(T_{S,j}, P_{{\sf sen},j}\left|n_{k,j}\right.)$ with respect to $T_{S,j}$ for given $P_{{\sf sen},j}$ in Proposition 1 in technical report \cite{report}, whose details are omitted due to the space constraint.
In fact, we prove that there exists the optimal solution for $\left(T_{S,j}^*, P_{{\sf sen},j}^*\right)$ to maximize the throughput 
$\mathcal{NT}_j(T_{S,j}, P_{{\sf sen},j}\left|n_{k,j}\right.)$.
Therefore, we can determine the optimal values $\left(T_{S,j}, P_{{\sf sen},j}\right)$ by using bi-section search of $P_{{\sf sen},j}$ for given corresponding optimal $T_{S,j}$ on each channel 
with the corresponding number of contending SUs.

\subsubsection{Configuration for Channel Selection Probabilities}

We now solve problem \textbf{P3} by employing the polynomial optimization technique (for details of this technique, please see \cite{Lasserre01}).
Let us define the variables $\vec{X} = \left\{X_1, \ldots, X_M, X_{M+1}, \ldots, X_{M+\left|\Omega\right|}\right\}$ as follows:  
\beqn
X_i = \left\{ {\begin{array}{*{20}{l}}
   p_i^{\sf sec} & \mbox{if } 1 \leq i \leq M  \\
   \prod_{j=1}^M 
	\! \left(p_j^{\sf sec}\right)^{n_{k,j}} & \mbox{if } 1\leq k = i - M \leq \left|\Omega\right|  \\
\end{array}} \right.
\eeqn
Then, problem \textbf{P3} can be transformed into the linear program
\vspace{0.2cm}
\noindent
\textbf{P4:} 
\vspace{0.1cm}
\beqn
\label{EQN_Problem4}
\begin{array}{l}
 {\mathop {\max }\limits_{\vec{X}}} \quad {\mathcal{NT}} \left(\vec{X}\right) =  \sum_{k=1}^{\left|\Omega\right|} X_{M+k} \mathcal{B} \left(\omega_k\right) \\ 
 \mbox{s.t.}\,\,\,\, \quad 0 \leq X_i \leq 1, i \in \left\{1, \ldots, M+\left|\Omega\right|\right\} \\
   \quad \quad \quad \sum_{i=1}^M X_i = 1 \\
 \end{array}
\eeqn
where $\mathcal{B} \left(\omega_k\right)$ is the constant, which is given in (\ref{EQN_B_constant}).
Recall that $\mathcal{B} \left(\omega_k\right)$ can be determined from the optimal solution of Problem \textbf{P2} in step 1.
To solve problem \textbf{P4} in step 2, standard methods in \cite{Lasserre01} such as cutting-plane method, branch and bound, branch and cut,
 branch and price can be employed. We use the branch and bound method \cite{Sherali91} to solve this problem.

\section{Numerical Results}
\label{Results}

To obtain numerical results, we take key parameters for the MAC protocol from Table II in \cite{Tan11}. All other parameters are chosen as follows unless stated otherwise:
mini-slot is $\sigma = 20 {\mu} s$; sampling frequency for sensing is $f_s = 6$MHz; bandwidth of PU's QPSK signal is 
$6$MHz; $\mathcal{\overline P}_d = 0.8$; the SNR of PU 
signals at SUs $\gamma_P = \frac{P_p}{N_0} = -20$dB; varying self-interference parameters $\zeta$ and $\xi$. Without loss of generality, the noise power is
 normalized to one; hence, the SU transmit power, $P_{\sf sen}$ becomes $P_{\sf sen} = SNR_s$; and $P_{\sf max} = 15$dB.
For one specific channel, we investigate the effect of self-interference on the throughput performance and the single-channel throughput performance versus 
SU transmit power $P_{\sf sen}$ and sensing time $T_S$ for different cases with varying self-interference parameters.
Detailed results are shown in the online technical report \cite{report} due to the space constraint.

\begin{figure}[!t]
\centering
\includegraphics[width=60mm]{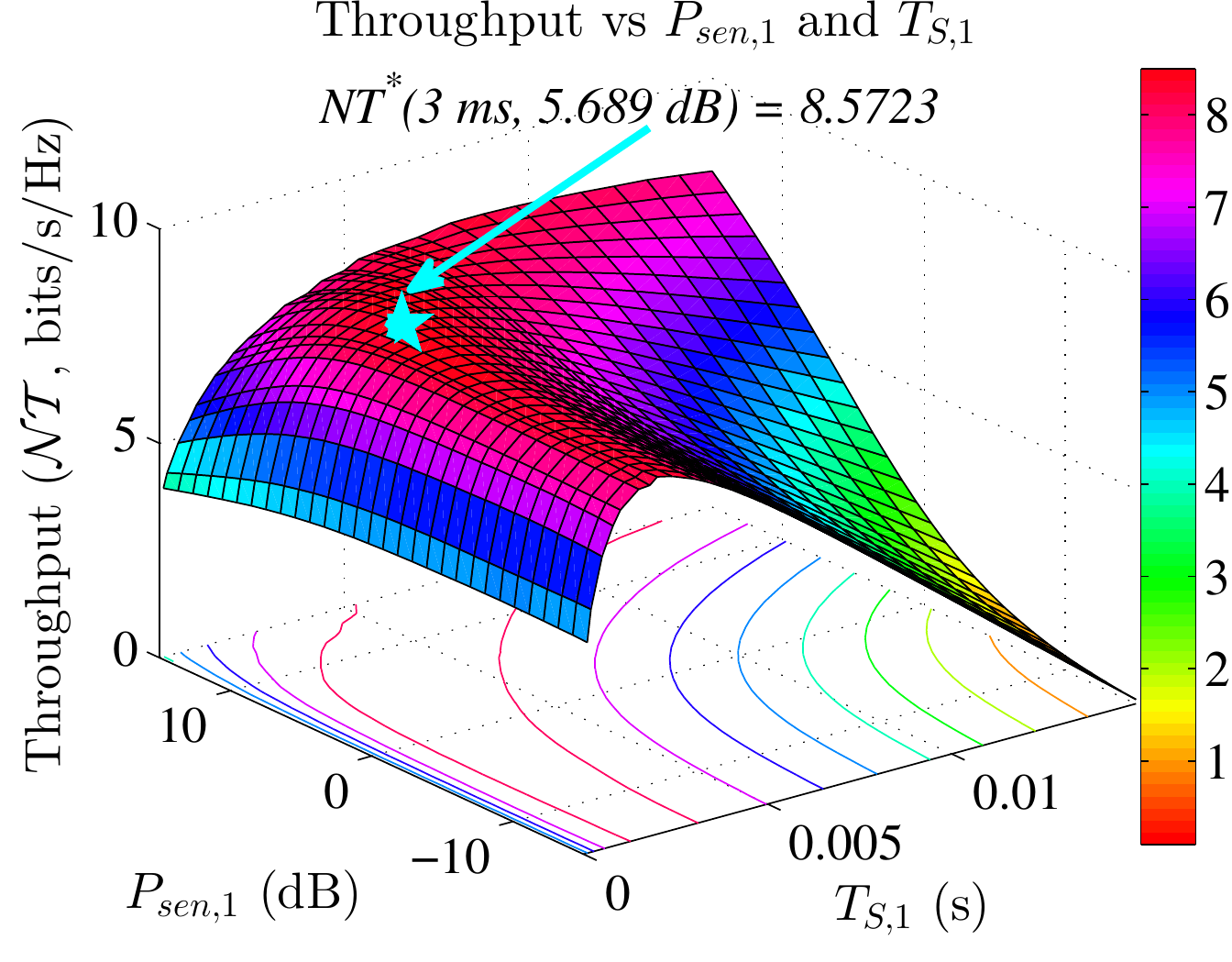}
\caption{Throughput versus SU transmit power $P_{\sf sen}$ and sensing time $T_S$ for $p = 0.0022$, $N = 50$, 
$\xi =0.95$, $\zeta = 0.2$ and  $P_{\sf dat} = 15$ dB.}
\label{P2_15dB_allchan_tauacid5025050}
\end{figure}

We now investigate the multichannel scenario where we consider a network consisting of $N = 50$ SUs and 3 channels with following parameter settings: 
${\bar \tau}_{\sf id}^1 = {\bar \tau}_{\sf id}^2 = {\bar \tau}_{\sf id}^3 = 1000$ ms, $\left\{{\bar \tau}_{\sf ac}^1, 
{\bar \tau}_{\sf ac}^2, {\bar \tau}_{\sf ac}^3  \right\}= {50,250, 50}$ ms.
Moreover, we set $p = 0.0022$ and $P_{\sf dat} = 15$ dB.
Fig.~\ref{P2_15dB_allchan_tauacid5025050} illustrates the throughput performance versus SU transmit power $P_{{\sf sen},1}$ and sensing time $T_{S,1}$ 
for channel 1 for the case with $\xi =0.95$ and $\zeta = 0.2$.
The optimal configuration of SU transmit power $P_{{\sf sen},1}^* = 5.689$ dB and sensing time $T_{S,1}^* = 3$ ms 
is shown to achieve the maximum throughput $\mathcal{NT}\left(T_{S,1}^*, P_{{\sf sen},1}^*\right) = 8.5723$, which is again indicated by a star symbol.

\begin{figure}[!t]
\centering
\includegraphics[width=60mm]{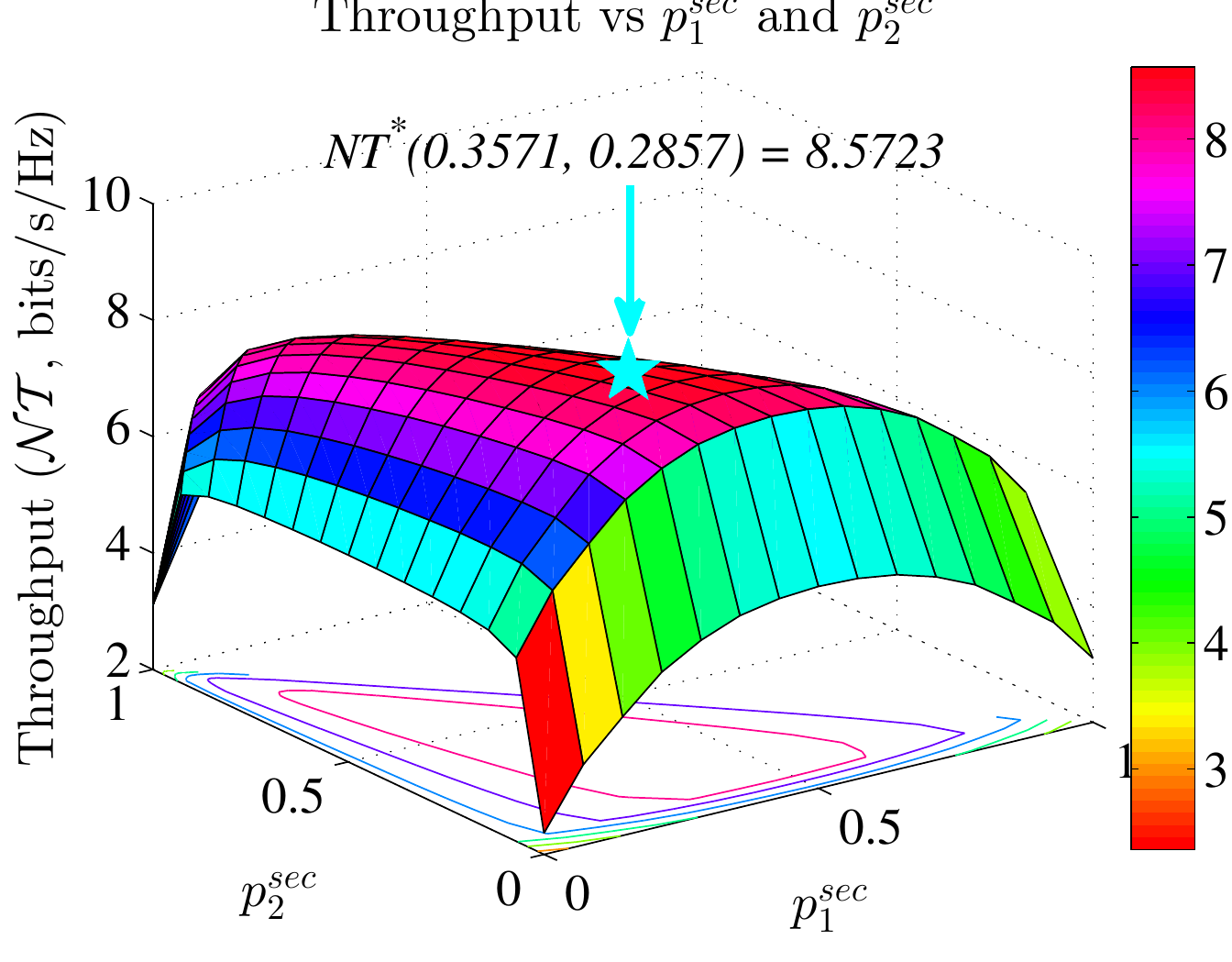}
\caption{Throughput versus SU transmit power $p_1^{\sf sec}$ and $p_2^{\sf sec}$ for $N = 50$, 
$\xi =0.95$, $\zeta = 0.2$ and $P_{\sf dat} = 15$ dB.}
\label{T_vs_psec1_psec2_max_tauac5025050}
\end{figure}

\begin{figure}[!t]
\centering
\includegraphics[width=60mm]{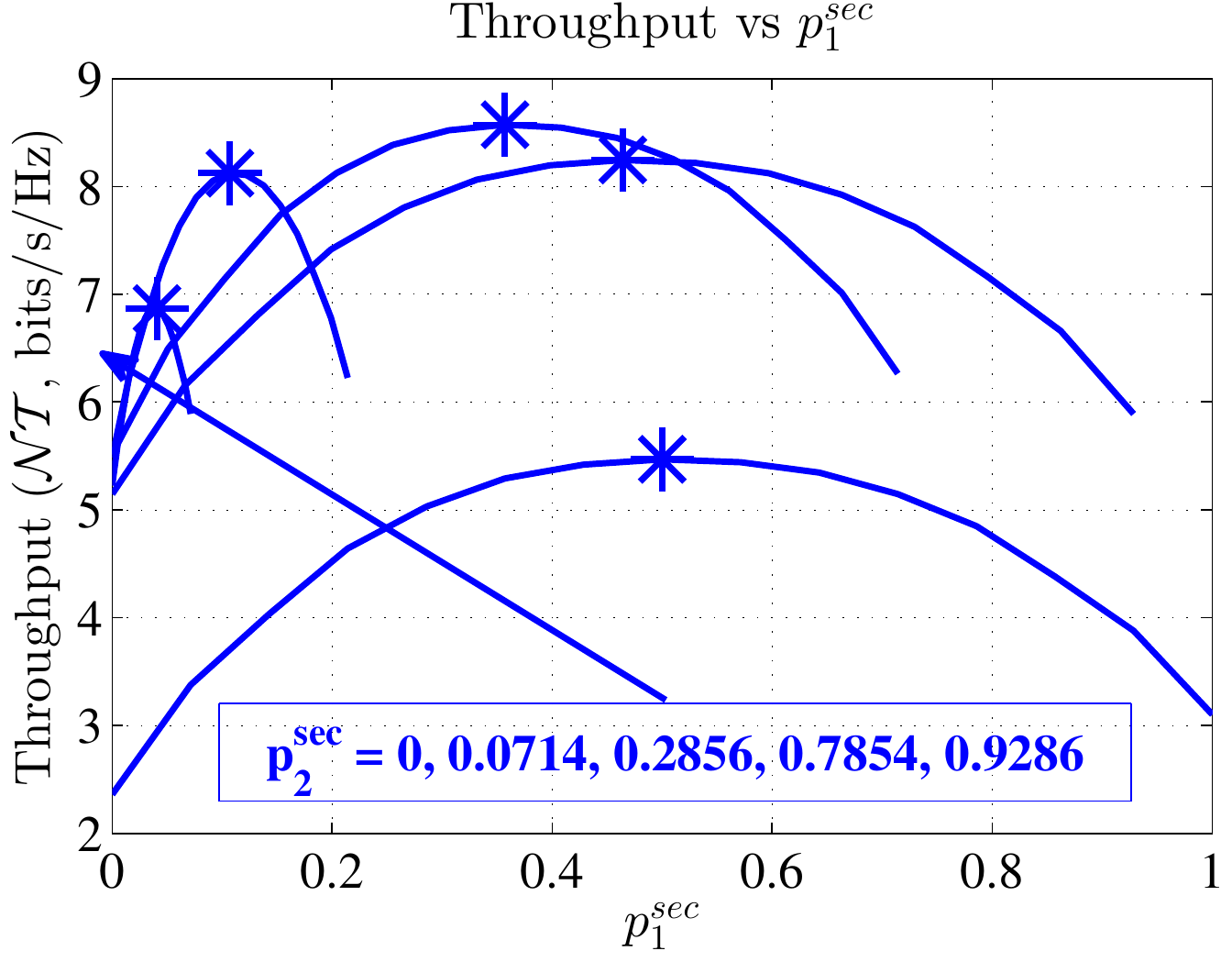}
\caption{Throughput versus SU transmit power $p_1^{\sf sec}$ and $p_2^{\sf sec}$ for $N = 50$, 
$\xi =0.95$, $\zeta = 0.2$ and $P_{\sf dat} = 15$ dB.}
\label{T_vs_psec1_max_tauac5025050}
\end{figure}

To investigate the impacts of  channel selection probabilities, we also consider the above network with following parameter
settings: ${\bar \tau}_{\sf id}^1 = {\bar \tau}_{\sf id}^2 = {\bar \tau}_{\sf id}^3 = 1000$ ms, $\left\{{\bar \tau}_{\sf ac}^1, {\bar \tau}_{\sf ac}^2, {\bar \tau}_{\sf ac}^3  \right\}= \left\{50,250, 50\right\}$ ms, $\xi =0.95$, $\zeta = 0.2$ and $P_{\sf dat} = 15$ dB.
Fig.~\ref{T_vs_psec1_psec2_max_tauac5025050} demonstrates the throughput performance versus channel selection  probabilities for 
channels 1 and 2 ($\left\{p_1^{\sf sec}, p_2^{\sf sec}\right\}$).
The optimal configuration of channel selection probabilities is  $p_1^{{\sf sec},*} = 0.3571$ and $p_2^{{\sf sec},*} = 0.2857$,
which is again indicated by a star symbol (i.e., the probability of channel selection for the last channel is $p_3^{{\sf sec},*} = 
1 - p_1^{{\sf sec},*} - p_2^{{\sf sec},*} = 0.3572$). 
We can observe that the SUs choose a more busy channel with lower channel selection probability at optimality, which
is quite intuitive.

To better observe the relationship of throughput vs channel selection probabilities, we show the throughput performance versus 
channel selection probabilities  for channels 1 and 2 ($\left\{p_1^{\sf sec}, p_2^{\sf sec}\right\}$) in Fig.~\ref{T_vs_psec1_max_tauac5025050}.
We set the network parameters as follows: ${\bar \tau}_{\sf id}^1 = {\bar \tau}_{\sf id}^2 = {\bar \tau}_{\sf id}^3 = 1000$ ms, $\left\{{\bar \tau}_{\sf ac}^1, {\bar \tau}_{\sf ac}^2, {\bar \tau}_{\sf ac}^3  \right\}= \left\{50,250, 50\right\}$ ms, $\xi = 0.95$, $\zeta = 0.2$, and $P_{\sf dat} = 15$ dB.
This figure shows that the throughput curve for each value of $p_2^{\sf sec}$ first increases to the maximum value (which is indicated by the asterisk) 
and then decreases as we increase $p_1^{\sf sec}$.

\begin{table} 
\centering
\caption{Throughput vs $\left(\tau_{\sf id}^1,\tau_{\sf id}^2\right)$ (MxN=2x20)}
\label{table1}
\begin{tabular}{|c|c|c|c|c|}
\hline 
\multicolumn{2}{|c|}{${\bar \tau}_{\sf id}^1$
ms}                  & 100  & 500 & 1000\tabularnewline
\hline
\hline 
Alg. 1 & $\mathcal{NT}$ & 3.7202  &  5.3001  &  5.9873  \tabularnewline
\hline
\hline 
Alg. 2 & $\mathcal{NT}$ & 3.6930  &  5.2752   & 5.9873  \tabularnewline
\hline 
\hline 
Proposed  & $p_1^{\sf sec}$ & 0.3218  & 0.4138 & 0.5 \tabularnewline
\cline{2-5} 
Alg. & $\mathcal{NT}$  & 4.0893  &  5.5870 &  5.9873 \tabularnewline
\cline{2-5} 
& $\Delta \mathcal{NT}_1 (\%)$  & 9.0260  &  5.1351  &      0 \tabularnewline
\cline{2-5} 
& $\Delta \mathcal{NT}_2 (\%)$  & 9.6924  &  5.5808   &      0 \tabularnewline
\hline
\end{tabular}
\end{table}

\begin{table*} 
\centering
\caption{Throughput vs $\left({\bar \tau}_{\sf id}^1,{\bar \tau}_{\sf id}^2\right)$ (MxN=3x30)}
\label{table2}
\begin{tabular}{|c|c|c|c|c|c|c|}
\hline 
\multicolumn{2}{|c|}{$\left({\bar \tau}_{\sf id}^1,{\bar \tau}_{\sf id}^2\right)$
ms}                  & (50,50)   & (500,50)   & (1000,50)   & (500,500)   & (1000,1000) \tabularnewline
\hline
\hline 
Alg. 1 & $\mathcal{NT}$ & 4.8111  &  6.6440  &  7.2786  &  8.3649  &  9.5121 \tabularnewline
\hline
\hline 
Alg. 2 & $\mathcal{NT}$ &  4.7611  &  6.5130   & 7.1366   & 8.2649   & 9.5121 \tabularnewline
\hline 
\hline 
 & $\left(p_1^{\sf sec},p_2^{\sf sec}\right)$ & (0.2511, 0.2511)  & (0.3888, 0.2012) & (0.4244, 0.1512) & (0.3011, 0.3011) & (0.3333, 0.3333)\tabularnewline
\cline{2-7} 
Proposed  & $\mathcal{NT}$  & 5.4150  &  7.3515  &  7.9532  &  8.8231  &  9.5121\tabularnewline
\cline{2-7} 
Alg. & $\Delta \mathcal{NT}_1 (\%)$  & 11.1524  &  9.6239  &  8.4821   & 5.1932     &    0 \tabularnewline
\cline{2-7} 
& $\Delta \mathcal{NT}_2 (\%)$  & 12.0757  &  11.4058  &  10.2676  &   6.3266   &       0  \tabularnewline
\hline
\end{tabular}
\end{table*}

We now consider the scenario with 20 SUs and 2 channels.
We set ${\bar \tau}_{\sf id}^2 = 1000$ ms, ${\bar \tau}_{\sf ac}^1 = {\bar \tau}_{\sf ac}^2 = 100$ ms, varying ${\bar \tau}_{\sf id}^1$, $\xi =1$, $\zeta = 0.3$,
 and $P_{\sf dat} = 15$ dB.
We would like to compare our proposed design with other two schemes (called Algs. 1 and 2) which do not optimize the channel selection probabilities. 
In Alg. 1, we use equal channel selection probabilities for different channels, i.e., $p_1^{\sf sec} = p_2^{\sf sec}$.
In Alg. 2, a fixed channel assignment is used, i.e, each channel is assigned to one corresponding set of SUs where each set has the same number of SUs.
For the channels with $\left\{{\bar \tau}_{\sf id}^i\right\} = \left\{100, 500, 1000\right\}$ ms, we obtain the corresponding throughput
values of $\left\{\mathcal{NT}_i\right\} = \left\{0.6992, 2.2815, 2.9937
\right\}$, respectively. So the total throughput is the sum of two throughput values for the two channels. 
Note that we also perform optimization of sensing and access parameters for each channel in calculating the throughput of Algs. 1 and 2. 
The results shown in Table~\ref{table1} demonstrate that our proposed algorithm outperforms Algs. 1 and 2.
Moreover, the throughput gains between our proposed algorithm and Algs. 1 and 2 (which are $\Delta \mathcal{NT}_1$ and $\Delta \mathcal{NT}_2$, respectively) 
are about 10\%, which is quite significant.

We now consider the scenario with 30 SUs and 3 channels.
We set ${\bar \tau}_{\sf id}^3 = 1000$ ms, ${\bar \tau}_{\sf ac}^1 = {\bar \tau}_{\sf ac}^2 = {\bar \tau}_{\sf ac}^3= 50$ ms, varying ${\bar \tau}_{\sf id}^1$ and ${\bar \tau}_{\sf id}^2$, $\xi =0.95$, $\zeta = 0.4$ and $P_{\sf dat} = 15$ dB.
We will also compare our proposed design with Algs. 1 and 2.
For the channels with $\left\{{\bar \tau}_{\sf id}^i\right\} = \left\{50, 500, 1000\right\}$ ms, we obtain
 the corresponding throughput values of $\left\{\mathcal{NT}_i\right\} = \left\{0.7952, 2.5471, 3.1707\right\}$.
Again, the total throughput for Alg. 2 is the sum of throughput values achieved for the three channels. 
The results summarized in Table~\ref{table2} show that our proposed design again outperforms Algs. 1 and 2.
Note that for the case of ${\bar \tau}_{\sf id}^1 = {\bar \tau}_{\sf id}^2 = {\bar \tau}_{\sf id}^3$, all algorithms 
achieve the same throughput because all the channels have the same statistics and we do not need to optimize the load balancing.

\begin{figure}[!t]
\centering
\includegraphics[width=60mm]{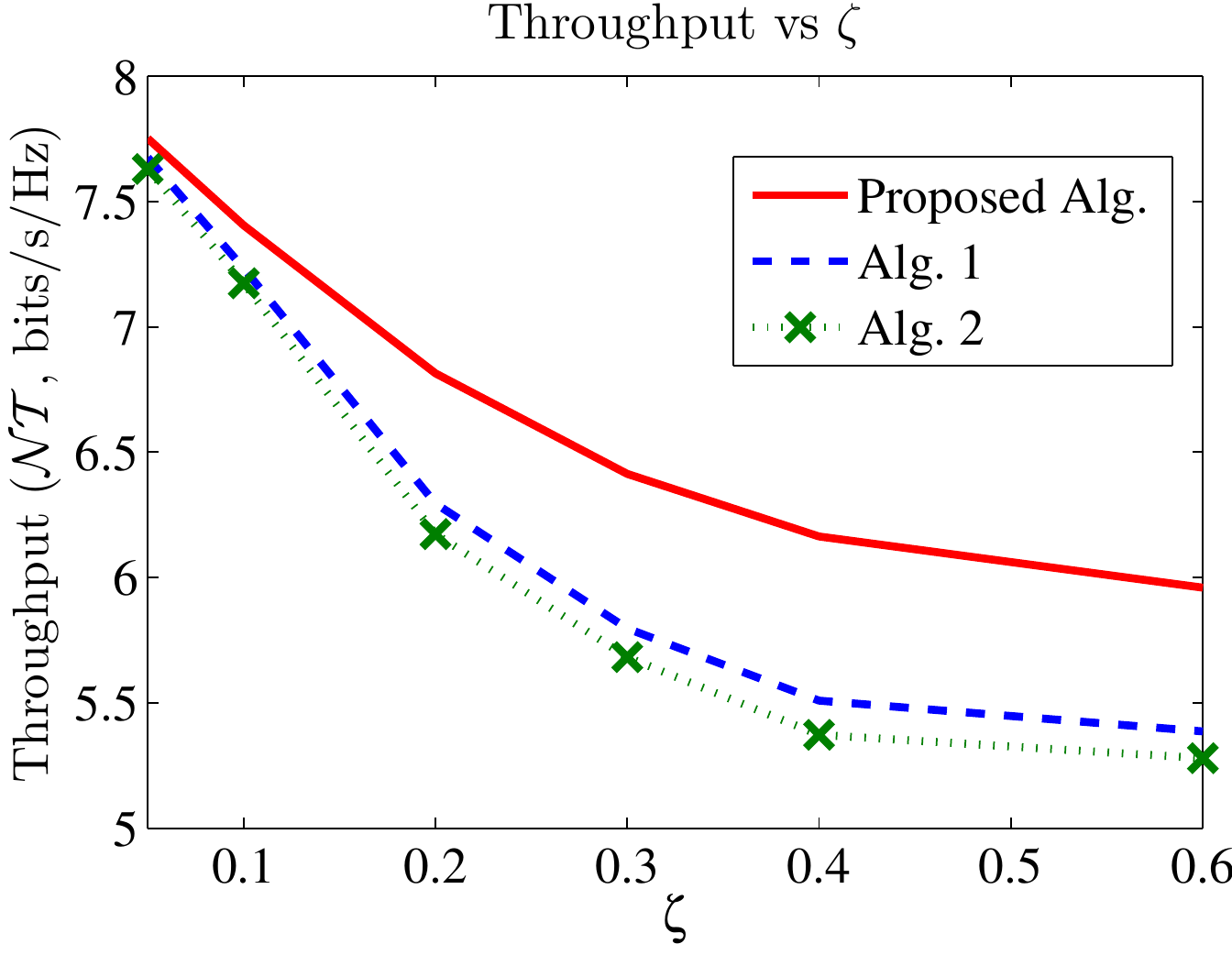}
\caption{Throughput versus $\zeta$ for $M \times N = 3 \times 30$, $\xi =0.95$, ${\bar \tau}_{\sf id}^j = 1000$ ms, $\left({\bar \tau}_{\sf ac}^1, {\bar \tau}_{\sf ac}^2, {\bar \tau}_{\sf ac}^3\right) = \left(50, 150, 500\right)$ ms and $P_{\sf dat} = 15$ dB.}
\label{Sel_interference_compare}
\end{figure}

In Fig.~\ref{Sel_interference_compare}, we compare our proposed design with Algs. 1 and 2 for varying self-interference parameters.
Here, we consider the network of 60 SUs and 3 channels where $\xi =0.95$, ${\bar \tau}_{\sf id}^j = 1000$ ms, $\left({\bar \tau}_{\sf ac}^1, {\bar \tau}_{\sf ac}^2, {\bar \tau}_{\sf ac}^3\right) = \left(50, 150, 500\right)$ ms and $P_{\sf dat} = 15$ dB.
Again, our proposed design leads to higher throughput than those under Algs. 1 and 2.
Moreover, the throughput gaps between our design and Algs. 1 and 2 become larger with lower self-interference cancellation quality.
These results confirm that optimal configuration of the MAC protocol and load balancing parameters 
is indeed important to achieve the largest throughput performance.

\section{Conclusion}
\label{conclusion} 

In this paper, we have proposed the MFDC--MAC protocol for CRNs, analyzed its throughput
performance, and studied its optimal parameter configuration. The design and analysis have taken into
account the FD communication capability and the self-interference of the FD transceiver.
In addition, we proposed the mechanism of channel selection to effectively balance the load among channels.
Finally, we have presented extensive numerical results to demonstrate the impacts
of self-interference and protocol parameters of sensing, access and load balancing strategies on the throughput performance.

\appendices

\section{False Alarm and Detection Probabilities}
\label{CAL_P_F_P_D}

We derive the detection and false alarm probabilities for FD sensing and two PU's state-changing events $h_{00}$ and $h_{01}$ in this appendix.
Here, we consider only one specific channel $j$, so we omit index $j$ in all the parameters for simplicity.
Assume that the transmitted signals from the PU and SU are circularly symmetric complex Gaussian (CSCG) signals while the noise at the secondary
receiver is independently and identically distributed CSCG $\mathcal{CN}\left( {0,{N_0}} \right)$ \cite{Liang08}. 
Under FD sensing, the  false alarm probability for event $h_{00}$ can be derived using the similar method as in \cite{Liang08},
which is given as
\beqn
\mathcal{P}_f^{00} = \mathcal{Q} \left[\left(\frac{\epsilon}{N_0+I(P_{\sf sen})}-1\right)\sqrt{f_sT_S}\right],
\eeqn
where $\mathcal{Q} \left(x\right) = \int_x^{+\infty} \exp \left(-t^2/2\right) dt$; $f_s$, $N_0$, $\epsilon$, $I(P_{\sf sen})$ are the sampling frequency, the noise power, the detection
threshold and the self-interference, respectively; $T_S$ is the FD sensing duration. 

The  detection probability for event $h_{01}$ is given as
\beqn
\mathcal{P}_d^{01} \!\! =  \mathcal{Q} \left( \!\!\frac{\left(\!\! \frac{\epsilon}{N_0+I(P_{\sf sen})}- \frac{T_S-t}{T_S}\gamma_{PS}-1\right) 
 \!\sqrt{f_sT_S}}{\sqrt{\frac{T_S-t}{T_S}\left(\gamma_{PS}+1\right)^2+\frac{t}{T_S}}} \!\! \right), 
\eeqn
where $t$ is the interval from the beginning of the data phase to the instant when the PU changes its state,
 $\gamma_{PS} = \frac{P_p}{N_0+I(P_{\sf sen})}$ is the signal-to-interference-plus-noise ratio (SINR) of the PU's signal at the SU.

\bibliographystyle{IEEEtran}


\end{document}